\documentclass[journal=jacsat,manuscript=article]{achemso}
\usepackage[version=3]{mhchem} 
\usepackage{siunitx}

\author{Maria Carolina Volpato}
\affiliation[Unicamp]
{Applied Physics Department and Photonics Research Center, “Gleb Wataghin” Institute of Physics, Universidade Estadual de Campinas, 13083-859, Campinas, SP, Brazil.}
\affiliation[NIST]
{National Institute of Standards and Technology, Gaithersburg, Maryland, USA.}
\author{Kalebe B. Estevam}
\affiliation[UFRN]
{Physics Department, Federal University of Rio Grande do Norte, 59078-900, Natal, RN, Brazil}
\author{Marcelo I. Davanco}
\affiliation[NIST]
{National Institute of Standards and Technology, Gaithersburg, Maryland, USA.}
\author{Pierre-Louis de Assis}
\email{plouis@unicamp.br}
\affiliation[Unicamp]
{Applied Physics Department and Photonics Research Center, “Gleb Wataghin” Institute of Physics, Universidade Estadual de Campinas, 13083-859, Campinas, SP, Brazil.}

\title{Broadband photonic structures to achieve high coupling efficiencies and Purcell factors with dark and interlayer excitons in 2D materials}

\begin{document}

\begin{abstract}
Horizontal slot waveguides are planar photonic structures with a guided mode which is strongly polarized in the out-of-plane direction and tightly confined in a sub-wavelength region of lower refractive index. We show through FDTD simulations that this mode can lead to coupling efficiencies $\beta>80\%$ and Purcell factors $F_P>10$  for some types of dark intralayer excitons in transition metal dichalcogenide (TMD) monolayers---more aplty named ``gray excitons'', as their out-of-plane dipole does couple to adequately polarized light---as well as interlayer excitons in TMD heterostructures. These figures indicate a path to the strong coupling regime for gray and interlayer excitons, while bright excitons are poorly coupled to the slot mode and experience Purcell suppression for $\lambda>\SI{1}{\micro\meter}$. A significant hurdle towards strong coupling, however, is the low oscillator strengths of these two excitonic species. We use the Tavis-Cummings model to show that a horizontal-slot racetrack resonator can overcome this difficulty and reach a cooperativity $C>>1$, albeit sacrificing the broadband characteristic of waveguides.
\end{abstract}

\section{Introduction}

Two-dimensional (2D) semiconductors such as transition metal dichalcogenides (TMDs) have emerged as promising candidates for quantum photonics, due to their ability to host tightly bound excitons with narrow emission lines and operation at or near room temperature \cite{palacios2017large, parto2021defect, zhao2021site}. 

Excitons in TMD monolayers can be categorized according to their spin and momentum selection rules. Bright excitons are spin-allowed and couple efficiently to light, leading to strong photoluminescence \cite{wang2018colloquium}. In contrast, dark excitons are spin- or momentum-forbidden, resulting in much lower radiative decay rates. A subset of these, sometimes referred to as ``gray'' excitons (a terminology that will be adopted in the rest of this work) posses total angular momentum 0 \cite{abdurazakov2023formation}, and is typically investigated with the help of strong magnetic fields that allow them to couple with light impinging perpendicularly to the monolayer \cite{robert2017fine, wang2017plane,plankl2021subcycle}. Gray excitonic states typically have longer lifetimes than the bright ones. In addition to the intralayer excitons presented before, TMD heterobilayers can host interlayer excitons, where the electron and hole reside in different materials \cite{rivera2015observation, nagler2017interlayer, zhang2020twist}. They are also considered dark, as the transition is indirect. 

Both gray and interlayer excitons possess out-of-plane dipole moments, making them weakly radiative and challenging to access with usual microscopy techniques. To measure these excitons, it is essential to improve their interaction with observable optical modes. Horizontal slot waveguides---planar photonic structures which support strongly confined modes polarized perpendicularly to the plane of the sample\cite{almeida2004guiding}---can accomplish this,reaching high coupling efficiencies ($\beta$) and Purcell factors (F$_P$) \cite{quan2009broadband,bisschop2015broadband}—leading to an improvement of emission rates by at least one order of magnitude, mitigating spurious effects from the environment and compensating for the low oscillator strength of these out-of-plane excitons.

In this work, we use finite-difference time-domain (FDTD) simulations to analyze the coupling efficiency and Purcell enhancement achievable in horizontal slot waveguides. In the Methods section, we detail our exploration of the parameter space for horizontal slot waveguide designs, aimed at optimizing its coupling to gray and interlayer excitons in TMDs. The results of each step are then presented and discussed. As optimal parameters are defined, we proceed to discuss how these devices might open a path for the study of our target excitonic species in a cavity QED context, and evaluate the experimental feasibility of this approach.

\section{Methods}
\begin{figure}[h!]
    \centering
    \includegraphics[trim={2cm 3cm 0cm 4cm},clip,width=0.8\linewidth]{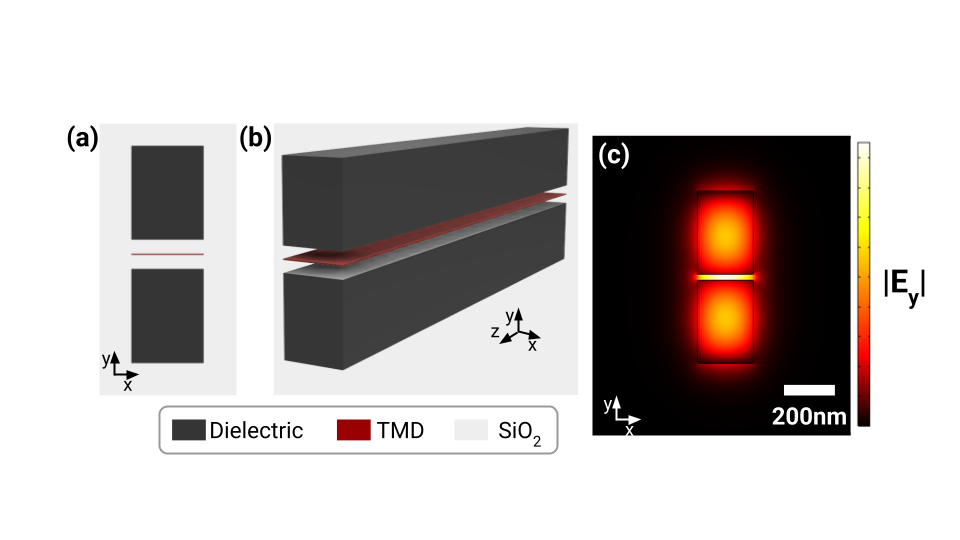}
    \caption{Horizontal slot waveguide geometry (a) and (b) with an embedded 2D material shown in red. This structures supports a mode which is polarized perpendicular to the dielectric interfaces and presents a high degree of sub-wavelength confinement, as seen in (c). Axes are oriented so that \textbf{x} is parallel to the TMD plane, \textbf{y} is parallel to the polarization of the slot guided mode, and \textbf{z} is the propagation direction for all guided modes.}
    \label{fig:slot}
\end{figure}

We modeled horizontal slot waveguides composed of a thin, low-index region composed of SiO$_2$ and embedding a 2D material monolayer, sandwiched between high-index claddings, as shown in Fig. \ref{fig:slot}(a-b). Fig.\ref{fig:slot}(c) shows that the fundamental guided mode in this geometry is strongly confined within the slot and polarized perpendicular to the interfaces, and to the plane of the 2D material. 

The emission wavelengths used in our FDTD simulations were chosen based on reported photoluminescence from gray and interlayer excitons in various TMD systems. For instance, MoS$_2$ and WSe$_2$ exhibit gray exciton emission in the visible to near-infrared range \cite{robert2020measurement, splendiani2010emerging, zhang2017magnetic, zhang2015experimental, terrones2014new}, while MoTe$_2$ emits in the near-infrared \cite{yang2015robust}. Interlayer excitons in heterostructures such as WSe$_2$/MoS$_2$ and WS$_2$/MoS$_2$ also span the near-infrared \cite{karni2019infrared, zhang2020twist}, and MoTe$_2$/MoS$_2$ stacks have been shown to emit up to the C-band telecom window \cite{zhang2016interlayer, ju2024infrared}. In order to cover the full range of reported excitonic emission across these materials, we ran simulations in three representative wavelength bands: \SIrange{640}{800}{\nm}, \SIrange{1050}{1150}{\nm}, and \SIrange{1500}{1600}{\nm} .

We started with FDTD simulations to determine which of the high-index dielectric platforms most commonly used in integrated photonics would yield the highest $\beta$ and F$_P$. We modeled waveguides using material parameters for slicon (Si), silicon nitride (SiNx), gallium phosphide (GaP), and lithium niobate on insulator (LN). The waveguide geometry, including total width and height, as well as the thickness of the SiO$_2$ layer, was independently optimized for each material in each wavelength band, using as a figure of merit for optimization the value of $\beta$ calculated between an out-of-plane (\textbf{y}-oriented) dipole and the slot mode. 

Dipole emitters oriented along the \textbf{x} (in-plane transverse), \textbf{y} (out-of-plane), and \textbf{z} (in-plane longitudinal) directions were placed at different positions within the slot to evaluate the spatial and polarization dependence of coupling efficiency and Purcell enhancement.

\section{Results}

\begin{figure}[h!]
    \centering
    \includegraphics[trim={6cm 0 6cm 0}, clip,height=0.5\linewidth]{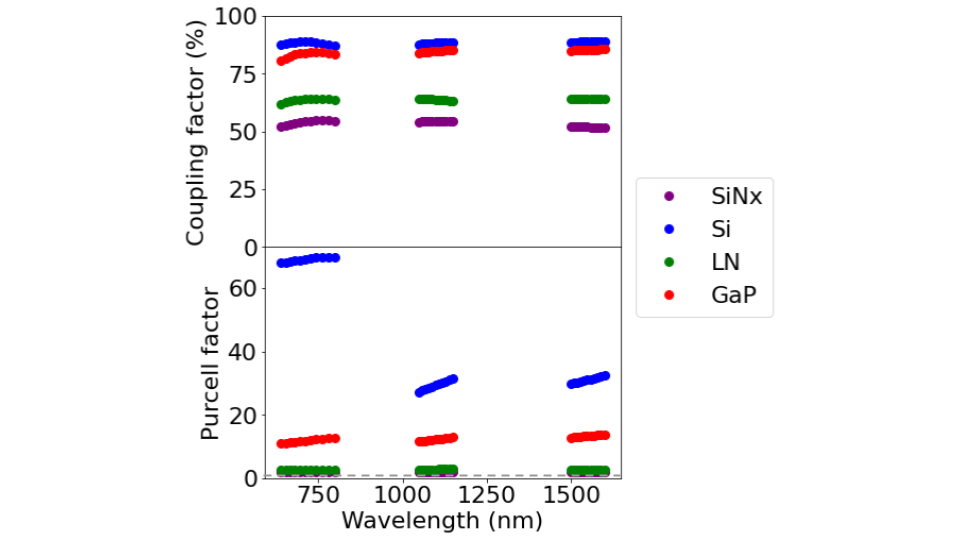}
    \caption{Effect of the choice of dielectric on the maximal $\beta$ and F$_P$ for dipoles oriented in the \textbf{y} direction (out of plane for a 2D material assumed to be at the center of the waveguide). Data shows coefficients for the optimal geometry for each wavelength. The dashed line indicates F$_P=1$. Values below it indicate emission suppression, instead of enhancement.}
    \label{fig:materials}
\end{figure}

Fig. \ref{fig:materials} shows $\beta$ and F$_P$ for SiNx, Si, LN and GaP. For each material, the waveguide geometry was optimized for each of the wavelength bands that we defined in the previous section. We found $\beta>$\SI{50}{\percent} for all materials. While Si structures were found with expected F$_P>70$, the average propagation losses of \SI{0.3}{\dB/\centi\meter} in the first and second bands, due to absorption, mean that they are not suitable for use in quantum applications. We found that GaP offers a better trade-off between field confinement and propagation loss for all bands, and we used it as the high-index dielectric for all other simulations.

\begin{figure}[!h]
    \centering
    \includegraphics[trim={4.5cm 0 4.5cm 0}, clip,height=0.5\linewidth]{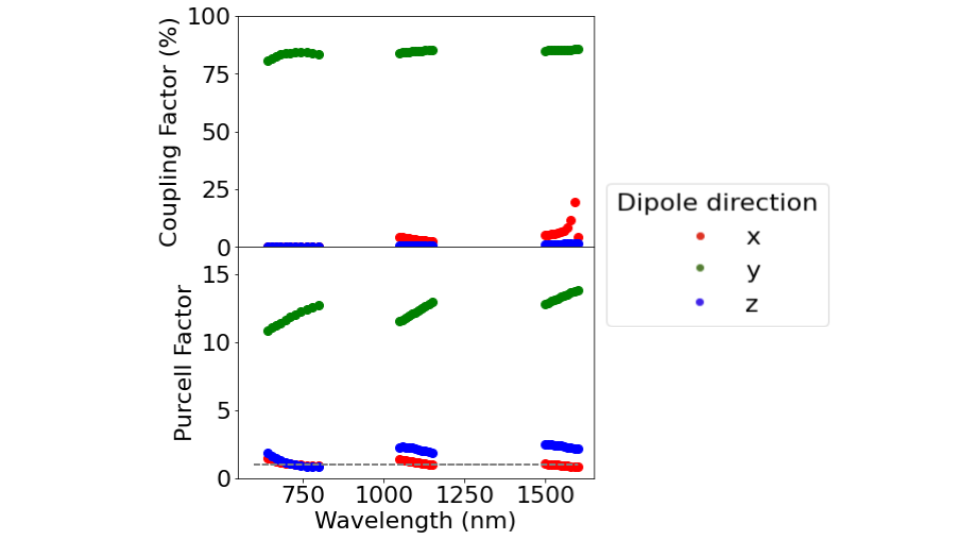}
    \caption{Dependence of $\beta$ and F$_P$ for dipoles oriented along \textbf{x}, \textbf{y} (already shown in Fig. \ref{fig:materials}, and \textbf{z}. We consider the dipoles to be at the center of the GaP slot. Data considers a constant (optimal) geometry for each band of wavelengths. The dashed line indicates F$_P=1$. Values below it indicate emission suppression, instead of enhancement.}
    \label{fig:directions}
\end{figure}

We proceeded to study how dipoles oriented along the three cartesian axes would couple to the slot mode, and what would F$_P$ be for each orientation. Fig. \ref{fig:directions} presents the coupling and Purcell factors for dipoles oriented along the \textbf{x}, \textbf{y}, and \textbf{z} directions in the GaP slot waveguide. As expected, the \textbf{y} dipoles showed a high $\beta$, above \SI{80}{\percent}, while the \textbf{x} and \textbf{z} dipoles (bright intralayer excitons in 2D materials) have negligible $\beta$ in the first and second bands under study. For the third band (telecom C band) the \textbf{x} dipole does present a more significant coupling, but is Purcell inhibited ($F_P<1$). It is important to note that the in-plane dipoles can couple to other modes of the waveguide, which are not under study in this work. This can allow bright excitons to be pumped by a mode orthogonal to the one under study, relax to the dark state, and then efficiently emit in the slot mode.

Finally, we report the result of the simulations done to evaluate $\beta$ and F$_P$ for emitters in other horizontal positions inside the slot waveguide, as the 2D material inside the slot is an extended system. Simulations were performed on different positions of the dipole in relation to the center of the waveguide, and we report results in terms of relative displacement, so that 0 represents a dipole centered on the optical axis and 1 represents a dipole at the edge of the waveguide.

\begin{figure}[h!]
    \centering
    \includegraphics[trim={4.8cm 0 4.8cm 0}, clip,height=0.5\linewidth]{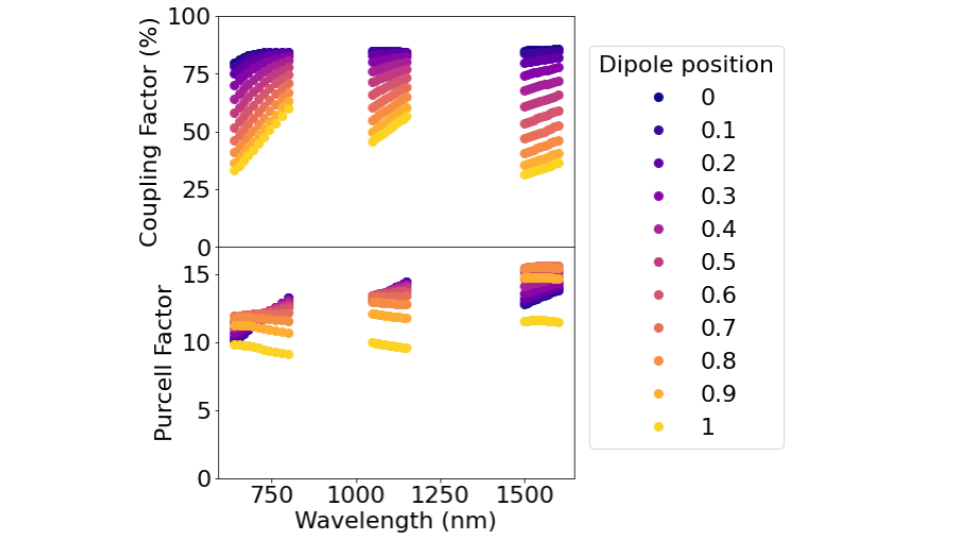}
    \caption{Effect of displacements from the optical axis on $\beta$ and F$_P$ for y-oriented dipoles in a GaP waveguide. Position 0 corresponds to the center of the slot (already shown in Fig. \ref{fig:materials}) and 1 to the edge of the waveguide. Data considers a constant (optimal) geometry for each band of wavelengths.}
    \label{fig:positions}
\end{figure}

It can be seen in Fig. \ref{fig:positions} that the highest values of $\beta$ occur when the dipole is positioned at the center of the slot, with performance gradually decreasing toward the waveguide edges. For F$_P$, however, in the first and third bands it is possible to see that higher values occur when the dipole is approximately at $x=\pm0.45w$, where $w$ is the full width of the waveguide. We attribute this to edge effects, and intend to investigate them further in a follow-up study. It is also noticeable that F$_P$ becomes less wavelength dependent as the dipole moves away from the optical axis, while $\beta$ has the opposite behavior.

\subsection{Cavity QED}

Dark and interlayer excitons exhibit lower oscillator strengths compared to bright excitons \cite{wang2017plane, guilhon2019out}, which can hinder the achievement of strong light–matter coupling. To overcome this limitation, a racetrack or ring resonator \cite{blaha2022beyond} can be fabricated, based on a slot waveguide, to provide resonant enhancement.

In addition to exploring the effect of strong coupling on a gas of excitons, it is also interesting to consider the possibility of having excitons bound to defects, acting as artificial atoms in this cavity. Localized single-photon emitters in 2D materials can be activated through strain engineering and defects \cite{parto2021defect}, so such a system could be fabricated without relying on naturally occurring emitters, which are randomly distributed. When modeling these systems, the Jaynes–Cummings (JC) and Tavis-Cummings (TC) formalism can be applicable \cite{jaynes2005comparison}. Within this context, the cooperativity can be defined as \cite{blaha2022beyond}.

\begin{equation}
C = \frac{g_1 ^2}{2\,\kappa_0,\gamma_l}, 
\label{eq1}
\end{equation}
where $g_1$ is the vacuum Rabi frequency of the coupled emitter-resonator system, $\kappa_0$ is the cavities field decay rate and $\gamma_l$ is the spontaneous emission rate of the emitter into all free-space modes. For a ring resonator, we can rewrite the Eq. \ref{eq1} as
\begin{equation}
C = \frac{\beta \nu_{\mathrm{FSR}}}{\kappa_0} (1 + F_P), 
\end{equation}
and 
\begin{equation}
\kappa_0 = \frac{2\pi c}{\lambda_0 \, Q_0}
\end{equation}
where $ \nu_{\mathrm{FSR}}$ is the free spectral range of the resonator, $c$ is the speed of light, $\lambda_0$ is the wavelength of the out-of-plane exciton and $Q_0$ is the intrinsic quality factor of the resonator.

For a typical cavity of $ \nu_{\mathrm{FSR}}=$\SI{500}{GHz}, $\lambda_0=$\SI{750}{nm} and $Q_0=10^4$, we estimate the cooperativity of $C\approx41\gg1$, for our results of $\beta$ and $F_p$, showing that is possible to achieve strong coupling in this system. We are also investigating the possibility of attaining other regimes with ensembles of localized emitters, as predicted by the TC formalism, such as superstrong coupling, if $g_N$, the collective Rabi frequency, exceeds $\nu_{\mathrm{FSR}}$.

\section{Conclusion}


We have shown that the simultaneously high coupling efficiency and Purcell factors available in slot waveguides can open a new way of exploring the Physics of such systems, as well as interlayer excitons in 2D heterostructures. While the small oscillator strength of these excitons presents a challenge to achieving strong coupling, we are currently investigating if resonant structures based on the proposed slot waveguides can be designed to overcome this difficulty. Finally, strain and defect engineering may enable the localization of individual dark excitons at the slot center, offering a scalable path to broadband integrated quantum photonic platforms based on 2D materials, as well as experiments in different cavity QED regimes.

\subsection{Data Availability}
All data that supports the findings of this study are available upon request.

\begin{acknowledgement}
The authors would like to thank Bruno R. Carvalho, Yara Gobato, Marcelo França Santos, Dario Gerace, and Maxime Richard for the fruitful discussions. This work was supported by the Air Force Office of Scientific Research (AFOSR) through Grant No. FA9550-20-1-0002.

\end{acknowledgement}

\bibliography{biblio}

\providecommand{\latin}[1]{#1}
\makeatletter
\providecommand{\doi}
  {\begingroup\let\do\@makeother\dospecials
  \catcode`\{=1 \catcode`\}=2 \doi@aux}
\providecommand{\doi@aux}[1]{\endgroup\texttt{#1}}
\makeatother
\providecommand*\mcitethebibliography{\thebibliography}
\csname @ifundefined\endcsname{endmcitethebibliography}  {\let\endmcitethebibliography\endthebibliography}{}
\begin{mcitethebibliography}{25}
\providecommand*\natexlab[1]{#1}
\providecommand*\mciteSetBstSublistMode[1]{}
\providecommand*\mciteSetBstMaxWidthForm[2]{}
\providecommand*\mciteBstWouldAddEndPuncttrue
  {\def\EndOfBibitem{\unskip.}}
\providecommand*\mciteBstWouldAddEndPunctfalse
  {\let\EndOfBibitem\relax}
\providecommand*\mciteSetBstMidEndSepPunct[3]{}
\providecommand*\mciteSetBstSublistLabelBeginEnd[3]{}
\providecommand*\EndOfBibitem{}
\mciteSetBstSublistMode{f}
\mciteSetBstMaxWidthForm{subitem}{(\alph{mcitesubitemcount})}
\mciteSetBstSublistLabelBeginEnd
  {\mcitemaxwidthsubitemform\space}
  {\relax}
  {\relax}

\bibitem[Palacios-Berraquero \latin{et~al.}(2017)Palacios-Berraquero, Kara, Montblanch, Barbone, Latawiec, Yoon, Ott, Loncar, Ferrari, and Atat{\"u}re]{palacios2017large}
Palacios-Berraquero,~C.; Kara,~D.~M.; Montblanch,~A. R.-P.; Barbone,~M.; Latawiec,~P.; Yoon,~D.; Ott,~A.~K.; Loncar,~M.; Ferrari,~A.~C.; Atat{\"u}re,~M. Large-scale quantum-emitter arrays in atomically thin semiconductors. \emph{Nature communications} \textbf{2017}, \emph{8}, 15093\relax
\mciteBstWouldAddEndPuncttrue
\mciteSetBstMidEndSepPunct{\mcitedefaultmidpunct}
{\mcitedefaultendpunct}{\mcitedefaultseppunct}\relax
\EndOfBibitem
\bibitem[Parto \latin{et~al.}(2021)Parto, Azzam, Banerjee, and Moody]{parto2021defect}
Parto,~K.; Azzam,~S.~I.; Banerjee,~K.; Moody,~G. Defect and strain engineering of monolayer WSe2 enables site-controlled single-photon emission up to 150 K. \emph{Nature communications} \textbf{2021}, \emph{12}, 3585\relax
\mciteBstWouldAddEndPuncttrue
\mciteSetBstMidEndSepPunct{\mcitedefaultmidpunct}
{\mcitedefaultendpunct}{\mcitedefaultseppunct}\relax
\EndOfBibitem
\bibitem[Zhao \latin{et~al.}(2021)Zhao, Pettes, Zheng, and Htoon]{zhao2021site}
Zhao,~H.; Pettes,~M.~T.; Zheng,~Y.; Htoon,~H. Site-controlled telecom-wavelength single-photon emitters in atomically-thin MoTe2. \emph{Nature communications} \textbf{2021}, \emph{12}, 6753\relax
\mciteBstWouldAddEndPuncttrue
\mciteSetBstMidEndSepPunct{\mcitedefaultmidpunct}
{\mcitedefaultendpunct}{\mcitedefaultseppunct}\relax
\EndOfBibitem
\bibitem[Wang \latin{et~al.}(2018)Wang, Chernikov, Glazov, Heinz, Marie, Amand, and Urbaszek]{wang2018colloquium}
Wang,~G.; Chernikov,~A.; Glazov,~M.~M.; Heinz,~T.~F.; Marie,~X.; Amand,~T.; Urbaszek,~B. Colloquium: Excitons in atomically thin transition metal dichalcogenides. \emph{Reviews of Modern Physics} \textbf{2018}, \emph{90}, 021001\relax
\mciteBstWouldAddEndPuncttrue
\mciteSetBstMidEndSepPunct{\mcitedefaultmidpunct}
{\mcitedefaultendpunct}{\mcitedefaultseppunct}\relax
\EndOfBibitem
\bibitem[Abdurazakov \latin{et~al.}(2023)Abdurazakov, Li, and Shim]{abdurazakov2023formation}
Abdurazakov,~O.; Li,~C.; Shim,~Y.-P. Formation of dark excitons in monolayer transition metal dichalcogenides by a vortex beam: Optical selection rules. \emph{Physical Review B} \textbf{2023}, \emph{108}, 125435\relax
\mciteBstWouldAddEndPuncttrue
\mciteSetBstMidEndSepPunct{\mcitedefaultmidpunct}
{\mcitedefaultendpunct}{\mcitedefaultseppunct}\relax
\EndOfBibitem
\bibitem[Robert \latin{et~al.}(2017)Robert, Amand, Cadiz, Lagarde, Courtade, Manca, Taniguchi, Watanabe, Urbaszek, and Marie]{robert2017fine}
Robert,~C.; Amand,~T.; Cadiz,~F.; Lagarde,~D.; Courtade,~E.; Manca,~M.; Taniguchi,~T.; Watanabe,~K.; Urbaszek,~B.; Marie,~X. Fine structure and lifetime of dark excitons in transition metal dichalcogenide monolayers. \emph{Physical review B} \textbf{2017}, \emph{96}, 155423\relax
\mciteBstWouldAddEndPuncttrue
\mciteSetBstMidEndSepPunct{\mcitedefaultmidpunct}
{\mcitedefaultendpunct}{\mcitedefaultseppunct}\relax
\EndOfBibitem
\bibitem[Wang \latin{et~al.}(2017)Wang, Robert, Glazov, Cadiz, Courtade, Amand, Lagarde, Taniguchi, Watanabe, Urbaszek, \latin{et~al.} others]{wang2017plane}
Wang,~G.; Robert,~C.; Glazov,~M.~M.; Cadiz,~F.; Courtade,~E.; Amand,~T.; Lagarde,~D.; Taniguchi,~T.; Watanabe,~K.; Urbaszek,~B.; others In-plane propagation of light in transition metal dichalcogenide monolayers: optical selection rules. \emph{Physical review letters} \textbf{2017}, \emph{119}, 047401\relax
\mciteBstWouldAddEndPuncttrue
\mciteSetBstMidEndSepPunct{\mcitedefaultmidpunct}
{\mcitedefaultendpunct}{\mcitedefaultseppunct}\relax
\EndOfBibitem
\bibitem[Plankl \latin{et~al.}(2021)Plankl, Faria~Junior, Mooshammer, Siday, Zizlsperger, Sandner, Schiegl, Maier, Huber, Gmitra, \latin{et~al.} others]{plankl2021subcycle}
Plankl,~M.; Faria~Junior,~P.~E.; Mooshammer,~F.; Siday,~T.; Zizlsperger,~M.; Sandner,~F.; Schiegl,~F.; Maier,~S.; Huber,~M.~A.; Gmitra,~M.; others Subcycle contact-free nanoscopy of ultrafast interlayer transport in atomically thin heterostructures. \emph{Nature Photonics} \textbf{2021}, \emph{15}, 594--600\relax
\mciteBstWouldAddEndPuncttrue
\mciteSetBstMidEndSepPunct{\mcitedefaultmidpunct}
{\mcitedefaultendpunct}{\mcitedefaultseppunct}\relax
\EndOfBibitem
\bibitem[Zhang \latin{et~al.}(2020)Zhang, Zhang, Wu, Wang, Gogna, Hou, Watanabe, Taniguchi, Kulkarni, Kuo, \latin{et~al.} others]{zhang2020twist}
Zhang,~L.; Zhang,~Z.; Wu,~F.; Wang,~D.; Gogna,~R.; Hou,~S.; Watanabe,~K.; Taniguchi,~T.; Kulkarni,~K.; Kuo,~T.; others Twist-angle dependence of moir{\'e} excitons in WS2/MoSe2 heterobilayers. \emph{Nature communications} \textbf{2020}, \emph{11}, 5888\relax
\mciteBstWouldAddEndPuncttrue
\mciteSetBstMidEndSepPunct{\mcitedefaultmidpunct}
{\mcitedefaultendpunct}{\mcitedefaultseppunct}\relax
\EndOfBibitem
\bibitem[Almeida \latin{et~al.}(2004)Almeida, Xu, Barrios, and Lipson]{almeida2004guiding}
Almeida,~V.~R.; Xu,~Q.; Barrios,~C.~A.; Lipson,~M. Guiding and confining light in void nanostructure. \emph{Optics letters} \textbf{2004}, \emph{29}, 1209--1211\relax
\mciteBstWouldAddEndPuncttrue
\mciteSetBstMidEndSepPunct{\mcitedefaultmidpunct}
{\mcitedefaultendpunct}{\mcitedefaultseppunct}\relax
\EndOfBibitem
\bibitem[Quan \latin{et~al.}(2009)Quan, Bulu, and Lon{\v{c}}ar]{quan2009broadband}
Quan,~Q.; Bulu,~I.; Lon{\v{c}}ar,~M. Broadband waveguide QED system on a chip. \emph{Physical Review A—Atomic, Molecular, and Optical Physics} \textbf{2009}, \emph{80}, 011810\relax
\mciteBstWouldAddEndPuncttrue
\mciteSetBstMidEndSepPunct{\mcitedefaultmidpunct}
{\mcitedefaultendpunct}{\mcitedefaultseppunct}\relax
\EndOfBibitem
\bibitem[Bisschop \latin{et~al.}(2015)Bisschop, Guille, Van~Thourhout, Hens, and Brainis]{bisschop2015broadband}
Bisschop,~S.; Guille,~A.; Van~Thourhout,~D.; Hens,~Z.; Brainis,~E. Broadband enhancement of single photon emission and polarization dependent coupling in silicon nitride waveguides. \emph{Optics express} \textbf{2015}, \emph{23}, 13713--13724\relax
\mciteBstWouldAddEndPuncttrue
\mciteSetBstMidEndSepPunct{\mcitedefaultmidpunct}
{\mcitedefaultendpunct}{\mcitedefaultseppunct}\relax
\EndOfBibitem
\bibitem[Robert \latin{et~al.}(2020)Robert, Han, Kapuscinski, Delhomme, Faugeras, Amand, Molas, Bartos, Watanabe, Taniguchi, \latin{et~al.} others]{robert2020measurement}
Robert,~C.; Han,~B.; Kapuscinski,~P.; Delhomme,~A.; Faugeras,~C.; Amand,~T.; Molas,~M.~R.; Bartos,~M.; Watanabe,~K.; Taniguchi,~T.; others Measurement of the spin-forbidden dark excitons in MoS2 and MoSe2 monolayers. \emph{Nature communications} \textbf{2020}, \emph{11}, 4037\relax
\mciteBstWouldAddEndPuncttrue
\mciteSetBstMidEndSepPunct{\mcitedefaultmidpunct}
{\mcitedefaultendpunct}{\mcitedefaultseppunct}\relax
\EndOfBibitem
\bibitem[Splendiani \latin{et~al.}(2010)Splendiani, Sun, Zhang, Li, Kim, Chim, Galli, and Wang]{splendiani2010emerging}
Splendiani,~A.; Sun,~L.; Zhang,~Y.; Li,~T.; Kim,~J.; Chim,~C.-Y.; Galli,~G.; Wang,~F. Emerging photoluminescence in monolayer MoS2. \emph{Nano letters} \textbf{2010}, \emph{10}, 1271--1275\relax
\mciteBstWouldAddEndPuncttrue
\mciteSetBstMidEndSepPunct{\mcitedefaultmidpunct}
{\mcitedefaultendpunct}{\mcitedefaultseppunct}\relax
\EndOfBibitem
\bibitem[Zhang \latin{et~al.}(2017)Zhang, Cao, Lu, Lin, Zhang, Wang, Li, Hone, Robinson, Smirnov, \latin{et~al.} others]{zhang2017magnetic}
Zhang,~X.-X.; Cao,~T.; Lu,~Z.; Lin,~Y.-C.; Zhang,~F.; Wang,~Y.; Li,~Z.; Hone,~J.~C.; Robinson,~J.~A.; Smirnov,~D.; others Magnetic brightening and control of dark excitons in monolayer WSe2. \emph{Nature nanotechnology} \textbf{2017}, \emph{12}, 883--888\relax
\mciteBstWouldAddEndPuncttrue
\mciteSetBstMidEndSepPunct{\mcitedefaultmidpunct}
{\mcitedefaultendpunct}{\mcitedefaultseppunct}\relax
\EndOfBibitem
\bibitem[Zhang \latin{et~al.}(2015)Zhang, You, Zhao, and Heinz]{zhang2015experimental}
Zhang,~X.-X.; You,~Y.; Zhao,~S. Y.~F.; Heinz,~T.~F. Experimental evidence for dark excitons in monolayer WSe 2. \emph{Physical review letters} \textbf{2015}, \emph{115}, 257403\relax
\mciteBstWouldAddEndPuncttrue
\mciteSetBstMidEndSepPunct{\mcitedefaultmidpunct}
{\mcitedefaultendpunct}{\mcitedefaultseppunct}\relax
\EndOfBibitem
\bibitem[Terrones \latin{et~al.}(2014)Terrones, Corro, Feng, Poumirol, Rhodes, Smirnov, Pradhan, Lin, Nguyen, El{\'\i}as, \latin{et~al.} others]{terrones2014new}
Terrones,~H.; Corro,~E.~D.; Feng,~S.; Poumirol,~J.; Rhodes,~D.; Smirnov,~D.; Pradhan,~N.; Lin,~Z.; Nguyen,~M.; El{\'\i}as,~A.; others New first order Raman-active modes in few layered transition metal dichalcogenides. \emph{Scientific reports} \textbf{2014}, \emph{4}, 4215\relax
\mciteBstWouldAddEndPuncttrue
\mciteSetBstMidEndSepPunct{\mcitedefaultmidpunct}
{\mcitedefaultendpunct}{\mcitedefaultseppunct}\relax
\EndOfBibitem
\bibitem[Yang \latin{et~al.}(2015)Yang, Lu, Myint, Pei, Macdonald, Zheng, and Lu]{yang2015robust}
Yang,~J.; Lu,~T.; Myint,~Y.~W.; Pei,~J.; Macdonald,~D.; Zheng,~J.-C.; Lu,~Y. Robust excitons and trions in monolayer MoTe2. \emph{ACS nano} \textbf{2015}, \emph{9}, 6603--6609\relax
\mciteBstWouldAddEndPuncttrue
\mciteSetBstMidEndSepPunct{\mcitedefaultmidpunct}
{\mcitedefaultendpunct}{\mcitedefaultseppunct}\relax
\EndOfBibitem
\bibitem[Karni \latin{et~al.}(2019)Karni, Barr{\'e}, Lau, Gillen, Ma, Kim, Watanabe, Taniguchi, Maultzsch, Barmak, \latin{et~al.} others]{karni2019infrared}
Karni,~O.; Barr{\'e},~E.; Lau,~S.~C.; Gillen,~R.; Ma,~E.~Y.; Kim,~B.; Watanabe,~K.; Taniguchi,~T.; Maultzsch,~J.; Barmak,~K.; others Infrared interlayer exciton emission in MoS 2/WSe 2 heterostructures. \emph{Physical review letters} \textbf{2019}, \emph{123}, 247402\relax
\mciteBstWouldAddEndPuncttrue
\mciteSetBstMidEndSepPunct{\mcitedefaultmidpunct}
{\mcitedefaultendpunct}{\mcitedefaultseppunct}\relax
\EndOfBibitem
\bibitem[Zhang \latin{et~al.}(2016)Zhang, Zhang, Cheng, Li, Wang, Wei, Zhou, Yu, Sun, Wang, \latin{et~al.} others]{zhang2016interlayer}
Zhang,~K.; Zhang,~T.; Cheng,~G.; Li,~T.; Wang,~S.; Wei,~W.; Zhou,~X.; Yu,~W.; Sun,~Y.; Wang,~P.; others Interlayer transition and infrared photodetection in atomically thin type-II MoTe2/MoS2 van der Waals heterostructures. \emph{ACS nano} \textbf{2016}, \emph{10}, 3852--3858\relax
\mciteBstWouldAddEndPuncttrue
\mciteSetBstMidEndSepPunct{\mcitedefaultmidpunct}
{\mcitedefaultendpunct}{\mcitedefaultseppunct}\relax
\EndOfBibitem
\bibitem[Ju \latin{et~al.}(2024)Ju, Cai, Jian, Hong, Sun, Wang, and Liu]{ju2024infrared}
Ju,~Q.; Cai,~Q.; Jian,~C.; Hong,~W.; Sun,~F.; Wang,~B.; Liu,~W. Infrared Interlayer Excitons in Twist-Free MoTe2/MoS2 Heterobilayers. \emph{Advanced Materials} \textbf{2024}, \emph{36}, 2404371\relax
\mciteBstWouldAddEndPuncttrue
\mciteSetBstMidEndSepPunct{\mcitedefaultmidpunct}
{\mcitedefaultendpunct}{\mcitedefaultseppunct}\relax
\EndOfBibitem
\bibitem[Guilhon \latin{et~al.}(2019)Guilhon, Marques, Teles, Palummo, Pulci, Botti, and Bechstedt]{guilhon2019out}
Guilhon,~I.; Marques,~M.; Teles,~L.; Palummo,~M.; Pulci,~O.; Botti,~S.; Bechstedt,~F. Out-of-plane excitons in two-dimensional crystals. \emph{Physical Review B} \textbf{2019}, \emph{99}, 161201\relax
\mciteBstWouldAddEndPuncttrue
\mciteSetBstMidEndSepPunct{\mcitedefaultmidpunct}
{\mcitedefaultendpunct}{\mcitedefaultseppunct}\relax
\EndOfBibitem
\bibitem[Blaha \latin{et~al.}(2022)Blaha, Johnson, Rauschenbeutel, and Volz]{blaha2022beyond}
Blaha,~M.; Johnson,~A.; Rauschenbeutel,~A.; Volz,~J. Beyond the Tavis-Cummings model: Revisiting cavity QED with ensembles of quantum emitters. \emph{Physical Review A} \textbf{2022}, \emph{105}, 013719\relax
\mciteBstWouldAddEndPuncttrue
\mciteSetBstMidEndSepPunct{\mcitedefaultmidpunct}
{\mcitedefaultendpunct}{\mcitedefaultseppunct}\relax
\EndOfBibitem
\bibitem[Jaynes and Cummings(2005)Jaynes, and Cummings]{jaynes2005comparison}
Jaynes,~E.~T.; Cummings,~F.~W. Comparison of quantum and semiclassical radiation theories with application to the beam maser. \emph{Proceedings of the IEEE} \textbf{2005}, \emph{51}, 89--109\relax
\mciteBstWouldAddEndPuncttrue
\mciteSetBstMidEndSepPunct{\mcitedefaultmidpunct}
{\mcitedefaultendpunct}{\mcitedefaultseppunct}\relax
\EndOfBibitem
\end{mcitethebibliography}

\end{document}